\documentstyle[12pt]{article}
\begin{document}


\begin{center}
{\Large{Coherent two pion photoproduction on $^{12}$C}}
\end{center}

\vspace{0.3cm}

\begin{center}
{\large{ S. S. Kamalov$^*$ and E. Oset}}
\end{center}

\begin{center}
{\small{\it 
Departamento de F\'{\i}sica Te\'orica and IFIC, Centro Mixto 
Universidad de Valencia - CSIC, 46100 Burjassot (Valencia) Spain.}}
\end{center}

\vspace{3cm}

\begin{abstract}
{\small{We develop the formalism for coherent two pion 
photoproduction in nuclei and perform actual calculations of 
cross sections for $\pi^-\pi^+$ and $\pi^0\pi^0$ photoproduction 
on $^{12}C$.  We find that due to the isospin symmetry  
the cross section for $\pi^0\pi^0$ production is very 
small and has a maximum when the pions propagate together. 
However, the kinematical region where the energies and polar angles 
of the two $\pi^0$ mesons are equal and their relative azimuthal 
angle $\phi=180^0$ is forbidden.  Conversely in the 
$\pi^-\pi^+$  production the pions prefer to have a relative 
azimuthal angle 180$^0$ and the production of the pions 
propagating  together is suppressed.  The dominant one-body 
mechanism in both channels is related to the excitation of the 
$\Delta$ isobar. Hence the reaction can serve as a source of 
information about $\Delta$'s properties in nucleus. 
We have found that the reaction is sensitive to effects of the pion and
$\Delta$ renormalization in the nuclear medium, similar to those  found 
in the coherent $(\gamma,\pi^0)$ reaction, but magnified because 
of the presence of the two pions. }}
\end{abstract}


\section{INTRODUCTION}

 Coherent reactions of particle production in nuclei are 
interesting processes in many senses. They involve spin sums in 
the amplitudes and hence act as filters of the interaction 
selecting only some of the ingredients in the elementary 
transition matrix element. For instance, coherent pion 
production in $(p,p')$ reactions in nuclei around the $\Delta$ 
region selects the spin-longitudinal part of the $NN\rightarrow 
N\Delta$ transition, while coherent photon production in the 
same region would select the spin-transverse part of the 
interaction. On the other hand, in the case of hadron production 
the results are sensitive to the final state interaction of the 
hadron with the nucleus, which allows one to obtain information 
on the hadron-nucleus optical potential.  This is particularly 
useful if the produced hadron  is a neutral one which cannot be 
used as a beam, like a $\pi^0$.

One of the traditional coherent production processes is the 
$(\gamma,\pi^0)$ reaction, which has been the subject of intense 
study \cite{Oset1,SW81,Koch1,Boffi,Aset,Lak,Carrasco}.  Here the 
results are rather sensitive to the way the pions and the 
$\Delta$ are renormalized in the nuclear medium and the reaction 
has been used as a test of the many-body theories which lead to 
the medium renormalization of such particles~\cite{Oset2}.

 Coherent two pion photoproduction in nuclei has not been 
studied either theoretically nor experimentally, but the advent 
of new facilities like the Mainz Microtron and TJNAF, where 
$(\gamma,\pi\pi)$ reactions on the proton are been 
investigated \cite{Brag,Hans}, opens reasonable expectations 
that such experiments could be undertaken soon. Some 
experimental attempts, so far inconclusive, have already been 
done \cite{Liang}. 

The reactions we are discussing are       
 
$$
\gamma + A(g.s.)\rightarrow A(g.s.) + \pi^- + \pi^+\,,\\
\eqno{(1a)}
$$

$$
\gamma + A(g.s.)\rightarrow A(g.s.) + \pi^0 + \pi^0\,,\\
\eqno{(1b)}
$$

\noindent
where $A(g.s.)$ is the nucleus in the ground state. In the 
present work we calculate for the first time cross sections for 
the coherent two pion photoproduction. The fact that a realistic 
model has been developed which leads to a good reproduction of 
the experimental cross sections for the $(\gamma,\pi^-\pi^+)$ 
and $(\gamma,\pi^0\pi^0)$ reactions on the 
proton \cite{Jose0,Jose1}, allows us to make predictions for the 
coherent $(\gamma,\pi\pi)$ reaction in those channels,  which 
should be of use  in the forthcoming experiments.

In addition to the reasons given above for the interest in the 
coherent $(\gamma,\pi^0)$ reaction, the coherent 
$(\gamma,\pi\pi)$ reaction introduces new elements. It provides 
us with a unique opportunity for the study of the two pion 
photoproduction on spin-isospin zero objects.  Here two pions 
are produced satisfying Bose-statistics and there are striking 
features which are tied to the isospin symmetry of the 
particles which would be interesting to observe.
 Eventual violations of this symmetry would also be 
interesting to the light of the much work devoted to this 
subject \cite{Wein,Beane,Cohen,Bern1}. 

On the other hand, the reaction acts as a filter of part of the interaction,
thus offering new information on the elementary ($\gamma, \pi \pi)$
amplitudes, in the absence of a complete set of polarization observables,
additional to the one obtained  in present unpolarized experiments.

In the present work we shall develop the formalism and make 
predictions for coherent $(\gamma,\pi^-\pi^+)$ and 
$(\gamma,\pi^0\pi^0)$ reactions in $^{12}C$. 
  
\section{Formalism}
 
\subsection{Symmetry properties and observables}

The most general amplitude of the coherent two pion photoproduction 
on the spin-zero nucleus in the $\gamma A$ center of mass frame and 
working in the Coulomb gauge ($\vec{k}\cdot\vec\epsilon=0, 
\,\,\epsilon^0=0$) is given by

$$
F^{(\lambda)}(\vec{q}_1,\vec{q}_2;\vec{k})=
F_{1}(\vec{q}_1,\vec{q}_2;\vec{k})\,\hat{q}_1\cdot
\vec{\epsilon}_{\lambda} +
F_{2}(\vec{q}_1,\vec{q}_2;\vec{k})\,\hat{q}_2\cdot
\vec{\epsilon}_{\lambda}\,, \\
\eqno{(2)}
$$

\noindent
where $\hat{q}_{1(2)}=\vec{q}_{1(2)}/q_{1(2)}$ is the unit 
vector for pion momentum, $\vec{k}$ is the photon momentum and 
$\vec{\epsilon}_{\lambda}$ with $\lambda=\pm 1$ is the photon 
polarization vector. In our study we will use a right-handed 
coordinate system in which the positive z-axis is along the 
photon momentum $\vec{k}$.

    The main symmetry properties of the $F_{1}$ and $F_{2}$ 
amplitudes are connected with the isospin (isoscalar+isovector) 
structure of the electromagnetic current. In the case of the 
isospin-zero target, an isoscalar photon $\gamma_S$ can produce 
two pions only in the state with isospin $T=0$ ( see Fig. 1a), 
which is necessarily symmetric. Bose-statistic requires that the 
orbital part of such system has to be symmetric too. Isovector 
photons $\gamma_V$ produce pions only in the state with isospin 
$T=1$ which requires an antisymmetrical orbital part. Thus, the 
amplitude for the coherent two pion photoproduction on 
isospin-zero targets can be divided into a symmetrical part 
$F^{(\lambda)}_S$, associated with scalar photons, and an 
antisymmetrical part $F^{(\lambda)}_V$, associated with vector 
photons:       

$$
F^{(\lambda)}(\vec{q}_1,\vec{q}_2;\vec{k})=
F^{(\lambda)}_S(\vec{q}_1,\vec{q}_2;\vec{k}) +
F^{(\lambda)}_V(\vec{q}_1,\vec{q}_2;\vec{k})\,. \\
\eqno{(3)}
$$

\noindent
The symmetry properties of the  $F^{(\lambda)}_S$ and  $F^{(\lambda)}_V$
amplitudes relative to the permutation $1 \leftrightarrow 2$ are the 
following

$$
F^{(\lambda)}_S(\vec{q}_1,\vec{q}_2;\vec{k})=
F^{(\lambda)}_S(\vec{q}_2,\vec{q}_1;\vec{k})\,,\quad\quad
F^{(\lambda)}_V(\vec{q}_1,\vec{q}_2;\vec{k})=
-F^{(\lambda)}_V(\vec{q}_2,\vec{q}_1;\vec{k})\,. \\
\eqno{(4)}
$$

\noindent
From Eq. (4) we can get the relations for the 
corresponding $F_{1}^{S,V}$ and $F_{2}^{S,V}$ amplitudes

$$
F_{1}^S(\vec{q}_1,\vec{q}_2;\vec{k})=
F_{2}^S(\vec{q}_2,\vec{q}_1;\vec{k})\,,\quad\quad
F_{1}^V(\vec{q}_1,\vec{q}_2;\vec{k})=
-F_{2}^V(\vec{q}_2,\vec{q}_1;\vec{k})\,. \\
\eqno{(5)}
$$

In the case of the coherent $(\gamma,\pi^0\pi^0)$ reaction the 
pions can be produced only in the symmetric $T=0$ state, but the 
$(\gamma,\pi^-\pi^+)$ channel contains contributions from both 
$T=0$ and $T=1$ pion isospin states. As we shall see below, this 
circumstance, combined with relations (5), will lead to an 
interesting physical phenomena.

Another peculiarity of the coherent two pion photoproduction is 
coming from  total angular momentum and parity conservation.  
The total angular momentum of the initial photon-nuclear system 
for a spin-zero nucleus is necessarily $J > 0$. Since parity 
and total momentum are conserved, the pions orbital momenta 
$\vec{l}_{1}$ and  $\vec{l}_{2}$  in the final state have to 
satisfy  the following selection rules

$$
\vec{l}_1+\vec{l}_2  =\vec{J} > 0;\qquad\qquad 
 l_1 + l_2 +J\, = \, even\,.  \\
\eqno{(6)}
$$

\noindent
Thus, the minimal total momentum  in the coherent double pion 
photoproduction on the spin-zero nuclei is $J_{min}=1$. 
Therefore, the pions can not be produced both in the $S$ state. 

   Additional symmetry properties are coming from the 
expressions for the differential cross section. As independent 
variables for the final three-particle state we shall take the 
pions solid angles $\Omega_1$ and $\Omega_2$ and the kinetic 
energy of one of the pions, for example, $T_1=\omega_1-m_{\pi}$. 
Then for the differential cross section we have 

$$
\frac{d\sigma}{dT_1 d\Omega_1 d\Omega_2} =
\frac{S_B}{(2\pi)^5}\,\frac{q_1 q_2 M_A^2}{8 E_{\gamma}\,E_A\, E'_A}\,
\frac{1}{\mid df/d\omega_2\mid}\frac{1}{2}\sum_{\lambda}
\mid F_{\lambda} \mid^2\,, \\
\eqno{(7)}
$$

\noindent
where $E_A\,(E'_A)$ is the nuclear energy in the initial (final) 
state, factor $S_B$=1 for $(\gamma,\pi^-\pi^+)$ and $S_B$=1/2 
for $(\gamma,\pi^0\pi^0)$ channels.  The recoil factor $\mid 
df/d\omega_2\mid$ in the phase space is given by

$$
\mid df/d\omega_2\mid = \mid 1 + \frac{\omega_2}{E'_A}
(1 + \frac{\vec{q}_1\cdot\vec{q}_2}{\vec{q}_2\!^2}) \mid\,. \\
\eqno{(8)}
$$

Since in the case of the $^{12}C$ target $\frac{\omega_2}{E'_A}<<1$
we neglect the nuclear recoil effects and take 

$$
\mid df/d\omega_2\mid\approx 1,\quad\quad E_A\approx E'_A\approx M_A,
\quad\quad \omega_2\approx E_{\gamma}-\omega_1\,. \\
\eqno{(9)}
$$

\noindent
Then after averaging over photon polarization $\lambda=\pm 1$ we 
can express the differential cross section in terms of the $F_1$ 
and $F_2$ amplitudes from Eq. (2) as

\begin{eqnarray}
\global\def\theequation{\arabic{equation}}
\setcounter{equation}{10}
\frac{d \sigma}{dT_1 d\Omega_1 d\Omega_2} & = &
\frac{S_B}{(2\pi)^5}\,\frac{q_1 q_2 }{16 E_{\gamma}} \,
\{ \mid F_1 \mid^2\sin^2\theta_1 +  \mid F_2 \mid^2\sin^2\theta_2 + 
\nonumber \\ & & 
 2\, Re(F_1 F_2^*) \sin \theta_1 \sin \theta_2\,\cos(\phi_1-\phi_2)\,\}
\,.  
\end{eqnarray}

Since $F_{1(2)}$ are scalar functions they can depend upon 
$\vec{k} \cdot \vec{q}_1,\,\vec{k} \cdot \vec{q}_2$ and 
$\vec{q}_1\cdot\vec{q}_2$, apart from the module of momenta. 
Hence $F_{1(2)}$ depends only on $\phi = \phi_1-\phi_2$ and so 
does the unpolarized cross section of Eq. (7). Note that, the 
photon asymmetry $\Sigma$ (for linear polarized photons) depends on 
$\phi_1$ and $\phi_2$ separately:

\begin{eqnarray}
\Sigma \, d \sigma\, & = & -\frac{S_B}{(2\pi)^5}\,\frac{q_1 q_2 }
{16 E_{\gamma}}\,\{\mid F_1 \mid^2\sin^2\theta_1\cos{2\phi_1} +
\mid F_2 \mid^2\sin^2\theta_2\cos{2\phi_2} + 
\nonumber \\ & &
2\,Re(F_1 F_2^*)\,\sin{\theta_1} \sin{\theta_2} \cos(\phi_1+\phi_2)\, \}
\,.  
\end{eqnarray}

\subsection{PWIA approach. Elementary amplitude}

The expressions considered above are general and they do not 
include information about the reaction mechanisms. To calculate 
the nuclear amplitudes $F_1$ and $F_2$ we will follow the 
traditional way by applying as a first step the Plane Wave 
Impulse Approximation (PWIA). In such framework we assume that: 
i) in the final state the effects from the interaction between 
particles can be neglected and ii) the nuclear amplitude is a 
coherent sum of the elementary amplitude $t^{(\lambda)}$ (which 
describe the process on the free nucleon) averaged over nucleon 
distribution in nuclei

$$
F^{(\lambda)}_{PWIA}(\vec{q}_1,\vec{q}_2;\vec{k})\equiv
V^{(\lambda)}(\vec{q}_1,\vec{q}_2;\vec{k})=
<\vec{q}_1,\vec{q}_2;0\mid \sum_{j=1}^{A}\,
t^{(\lambda)}(\vec{q}_1,\vec{q}_2;\vec{k},\vec{p}_j)\mid 0,\vec{k}>\,, \\
\eqno{(12)}
$$

\noindent
where $\mid 0>$ is the nuclear ground state, the $\mid 
 \vec{q}_1>,\, \mid \vec{q}_2>$ and $\mid \vec{k}>$ are pions 
 and photon plane waves.  Note that in the general case the 
elementary amplitude depends also upon the nucleon momentum 
$\vec{p}_j$ of the initial state.  We shall take this dependence 
into account approximately applying the well-known folding 
procedure (factorization approximation) \cite{Carrasco,Landau}

$$
\vec{p}_j\rightarrow \vec{p}_{eff}=-\frac{1}{2}(\vec{k}-
\vec{q}_1-\vec{q}_2)\,. \\
\eqno{(13)}
$$

In the coherent two pion photoproduction on the spin-zero 
nuclei only the non-spin-flip part of the elementary amplitude 
contributes. Using the factorization approximation (13) it can 
be reduced to an expression with the same structure as Eq. (2), 
i.e.

$$
t^{(\lambda)}(\vec{q}_1,\vec{q}_2;\vec{k})=
t_{1}(\vec{q}_1,\vec{q}_2;\vec{k})\,\hat{q}_1
\cdot\vec{\epsilon}_{\lambda} + t_{2}(\vec{q}_1,\vec{q}_2;\vec{k})\,
\hat{q}_2\cdot\vec{\epsilon}_{\lambda}\,. \\
\eqno{(14)}
$$

Since we are dealing with isospin-zero nuclear targets (with 
equal numbers of protons and neutrons) we need only the 
isoscalar  component of the elementary amplitude $t_{1(2)}$ (in 
the nucleon isospin space), which can be determined as

$$
t_{1(2)}=\frac{1}{2}\left[ t_{1(2)}^p + t_{1(2)}^n \right]\,, \\
\eqno{(15)}          
$$

\noindent
where $t_{1(2)}^p$ and $t_{1(2)}^n$ are the amplitudes for the 
proton and neutron channels, respectively. We obtain the final 
expression for the $F_{1(2)}$ amplitudes in the PWIA approach  
by substituting  Eq. (14) in  Eq. (12): 

$$
F_{1(2)}^{PWIA}(\vec{q}_1,\vec{q}_2;\vec{k})\equiv
V_{1(2)}(\vec{q}_1,\vec{q}_2;\vec{k})\,=\,
A\, F_A (Q)\,t_{1(2)}(\vec{q}_1,\vec{q}_2;\vec{k})\,, \\
\eqno{(16)}
$$        

\noindent
where $A$ is the number of nucleons and 
$\vec{Q}=\vec{k}-\vec{q}_1-\vec{q}_2$ is the transferred momentum,
which is given by

$$
Q^2=k^2+q_1^2+q_2^2-2k(q_1\cos\theta_1+q_2\cos\theta_2)+
2q_1q_2\cos\theta_{12}\,, \\
\eqno{(17a)}
$$

$$
\cos\theta_{12}=\cos\theta_1\cos\theta_2+\sin\theta_1\sin\theta_2\,
\cos(\phi_1-\phi_2)\,. \\
\eqno{(17b)}
$$

\noindent
The nuclear form factor $F_A(Q)$ (which is normalized to 1 at 
$Q$=0) is extracted from the well-known nuclear charge form 
factor using the relation $F_A^{ch.}(Q)=F_A(Q)f_p^{ch.}(Q)$ 
where $f_p^{ch.}$ is the proton charge form factor. 

One of the basic elements in the expression of Eq. (16) is the 
elementary amplitude $t_{1(2)}$ which is the spin-isospin 
independent part of the total amplitude describing the process 
on the free nucleons.  In the present paper we shall evaluate it 
using the recently developed model of Refs. \cite{Jose0,Jose1}.

In the Ref. \cite{Jose0} the model for the $\gamma 
p \rightarrow \pi^-\pi^+ p$ reaction was described. It was 
constructed from effective Lagrangians involving the coupling of 
photons and pions to nucleons and resonances. The 
$\Delta(1232),\,N^*(1440)$ and $N^*(1520)$ resonances were 
considered, as well as the coupling of $\rho$-mesons to the two 
pions system and 67 Feynmann diagrams were evaluated. In 
Ref. \cite{Jose1} the model was improved introducing the $d$-wave 
part of the $N^*(1520)\rightarrow\Delta\pi$ decay and other 
refinements.  At the same time the number of diagrams was 
reduced to 20, which are shown in Fig. 2, once the relevant 
terms for energies below $E_{\gamma}$ = 800 MeV were found in 
Ref. \cite{Jose0}.  Also the model was extended to the different 
isospin channels.  We take thus the model of Fig. 2. Details on 
 the Effective Lagrangians needed for their evaluation can be 
seen in Ref. \cite{Jose1} and further details on the explicit 
amplitudes in Ref. \cite{Jose}.

In the  $\gamma p\rightarrow\pi^-\pi^+ p$ amplitude the dominant 
terms are the  $\Delta N \pi \gamma$ Kroll-Ruderman term (i)
which appears after the minimal substitution in the $\Delta N\pi$ 
vertex, the pion pole term (j) and the $N^*(1520)\Delta\pi$ term (p) 
which interferes with the  $\Delta N \pi \gamma$ Kroll-Ruderman 
one leading to a peak in the cross section around 
$E_{\gamma}$ = 800  MeV. For the case of the $\gamma 
p\rightarrow\pi^0\pi^0 p$ reaction the  $\Delta N\pi\gamma$ 
Kroll-Ruderman and pion-pole terms vanish and in this case 
other $\Delta$ terms and the $N^*(1520) \Delta\pi$ term give the 
largest contribution.

The $\Delta N\pi\gamma$ Kroll-Ruderman term is of an isovector 
nature, it contributes to $F^{(\lambda)}_V$. The averaged 
amplitude over proton and neutron of the diagram (i), 
considering the permutation of the $\pi^+$ and $\pi^-$ in the 
two pion lines of the diagram, is given by  

$$
t^{(\lambda)}=e\left(\frac{2}{3}\frac{f^*}{m_{\pi}}\right)^2
\left[\frac{\vec{q}\,^*_{\pi^-}
\cdot\vec{\epsilon}_{\lambda}}{\sqrt{s_{\Delta_1}}-M_{\Delta}+
\frac{i}{2}\Gamma}\,-\, \frac{\vec{q}\,^*_{\pi^+}\cdot
\vec{\epsilon}_{\lambda}}{\sqrt{s_{\Delta_2}}-M_{\Delta}+
\frac{i}{2}\Gamma}\right]\,, \\
\eqno{(18)}
$$

\noindent
where $e$ is the proton charge, $f^*$ is the $\pi N \Delta$ coupling 
and $\vec{q}\,^*_{\pi}$ is the pion momentum in the $\Delta$ rest frame. 
In the nuclear calculations performed in the $\gamma A$ c.m. 
frame  we would express these momenta in terms
of those in the $\gamma \, A$ c.m. frame by means
of a boost \cite{Carrasco}. Furthermore

$$
s_{\Delta_1}=(p_f+p_{\pi^-})^2, \qquad \qquad 
s_{\Delta_2}=(p_f+p_{\pi^+})^2\,. \\
\eqno{(19)}
$$

We can see that Eq. (18) has the structure of the general 
amplitude of Eq. (2) and the amplitude is of the 
$F^{(\lambda)}_V$ type, fulfilling the second of the Eq. (4). 
This is a consequence of the $T_3^{\dagger}$ operator in the 
$\gamma\Delta N\pi$ vertex, where  $T_3^{\dagger}$ is the 
isospin 1/2 to 3/2 transition operator~\cite{Jose1}.

A trivial consequence is that, if 
$\vec{q}\,^*_{\pi^-}=\vec{q}\,^*_{\pi^+}$, the amplitude 
vanishes, which means that the dominant term in the elementary 
reaction gives null contribution when the $\pi^-,\pi^+$ are 
produced coherently and propagate together. In the case 
$\phi=\phi_1-\phi_2=180^0$ ( see Eq. (10)) the  $\Delta 
N\pi\gamma$ Kroll-Ruderman term (i) and other terms of isovector 
nature will contribute maximally.  Thus the strength of the 
cross section as a function of $\phi$ can be traced  to a 
varying weight of the different terms of the elementary 
$(\gamma,\pi\pi)$ amplitude, wich gives chances to learn about
the dynamics of the elementary process from
the experimental measurements of such cross sections.

\subsection{Final State Interaction (DWIA)}

From the study of the nuclear single pion photoproduction (see
for example Refs. \cite{SW81,Koch1,Boffi,Aset,Carrasco}) it is well 
known that the interaction of the pions with the residual 
nucleus ( Final State Interaction - FSI) is very important. At 
low energies the FSI increases the cross section and in the 
$\Delta$-resonance region reduces it (approximately in about a 
factor 2). In the coherent two pion photoproduction we expect 
that FSI effects can be more important.  The standard way to 
take into account FSI effects is to use the Distorted Wave 
Impulse Approximation (DWIA).  In the present paper we shall 
develop this method in the momentum space to allow for a more 
accurate treatment of the nonlocal nature of the elementary 
operator. In such approach the DWIA amplitude can be presented 
in the form 

$$ 
F^{(\lambda)}(\vec{q}_1,\vec{q}_2;\vec{k})= \int 
\Psi_{\vec{q}_1}(\vec{q}_1\!')\,\Psi_{\vec{q}_2}(\vec{q}_2\!')\, 
V^{(\lambda)}(\vec{q}_1\!',\vec{q}_2\!';\vec{k})\,d\vec{q}_1\!'\, 
d\vec{q}_2\!'\,.\\
\eqno{(20)}
$$

Note that here we assumed that there is no pion-pion 
interaction.  In the standard approach the pion distorted wave 
function $\Psi_{\vec{q}}(\vec{q}\,')$, which includes the 
effects of the pion-nuclear interaction, is a solution of the 
Klein-Gordon equation with a phenomenological optical potential. 
It can be expressed formally in terms of the  pion-nuclear 
scattering amplitude $F_{\pi A}$:

$$
\Psi_{\vec{q}}(\vec{q}\,')=\delta(\vec{q}-\vec{q}\,')-\frac{1}{(2\pi)^2}
\frac{F_{\pi A}(\vec{q},\vec{q}\,')}{{\cal M}(q')[ E(q)-E(q')+i\epsilon]}\,,
\eqno{(21)} \\
$$

\noindent
where $E(q)=\omega(q) + E_A(q)$ is the total energy of the 
pion-nuclear system and ${\cal M}(q)=\omega(q)E_A(q)/E(q)$ is the 
pion-nuclear relativistic reduced mass. 

In the present work the contributions from the FSI will be 
expressed in terms of the elastic pion-nuclear scattering 
amplitude, which was obtained as a solution of 
Lippmann-Schwinger equation with the phenomenological potential 
from Ref.~\cite{GKM}. After the substitution of Eq. (21) in Eq.  
(20) the final DWIA amplitude can be presented as a sum of the 
four terms

$$
F^{(\lambda)}(\vec{q}_1,\vec{q}_2;\vec{k})=
V^{(\lambda)}(\vec{q}_1,\vec{q}_2;\vec{k})+
D^{(\lambda)}_{1}(\vec{q}_1,\vec{q}_2;\vec{k})+
D^{(\lambda)}_{2}(\vec{q}_1,\vec{q}_2;\vec{k})+
D^{(\lambda)}_{12}(\vec{q}_1,\vec{q}_2;\vec{k})\,, \\
\eqno{(22)}
$$

\noindent
where

$$
D^{(\lambda)}_{1}(\vec{q}_1,\vec{q}_2;\vec{k})=
-\frac{1}{(2\pi)^2}\int \frac{d\vec{q}_1\!'}{{\cal M}(q'_1)}
\frac{F_{\pi A}(\vec{q}_1,\vec{q}_1\!')\,
V^{(\lambda)}(\vec{q}_1\!',\vec{q}_2;\vec{k})}{ E(q_1)-E(q'_1)+i\epsilon}\,,
\\ \eqno{(23a)}
$$

$$
D^{(\lambda)}_{2}(\vec{q}_1,\vec{q}_2;\vec{k})=
-\frac{1}{(2\pi)^2}\int \frac{d\vec{q}_2\!'}{{\cal M}(q'_2)}
\frac{F_{\pi A}(\vec{q}_2,\vec{q}_2\!')\,
V^{(\lambda)}(\vec{q}_1,\vec{q}_2\!';\vec{k})}{ E(q_2)-E(q'_2)+i\epsilon}\,,
\\
\eqno{(23b)}
$$

$$
D^{(\lambda)}_{12}(\vec{q}_1,\vec{q}_2;\vec{k})=
\frac{1}{(2\pi)^4}\int \frac{d\vec{q}_1\!'\,d\vec{q}_2\!'}{{\cal M}(q'_1)
{\cal M}(q'_2)}\frac{F_{\pi A}(\vec{q}_1,\vec{q}_1\!')
F_{\pi A}(\vec{q}_2,\vec{q}_2\!')\,V^{(\lambda)}(\vec{q}_1\!',
\vec{q}_2\!';\vec{k})}
{ [E(q_1)-E(q'_1)+i\epsilon]\,[E(q_2)-E(q'_2)+i\epsilon]}\,.
\\ \eqno{(23c)}
$$

\noindent
The first term is the PWIA amplitude which was considered above. 
The second and third terms ($D_{1}$ and $D_{2}$) describe the 
processes when only one pion interacts with the residual nuclei. 
The last term $D_{12}$ includes interaction of both pions.    

In order to facilitate the numerical analysis it is convenient 
to introduce partial wave amplitudes. For the pion-nuclear 
scattering amplitude we will use standard partial wave 
amplitudes defined as

$$
F_{\pi A}(\vec{q},\vec{q}\,')=4\pi\,\sum_{lm}{\cal F}_l(q,q')\,
Y_{lm}(\Omega_{\hat{q}})\,Y_{lm}^*(\Omega_{\hat{q}\,'})\,,
\eqno{(24)}
$$

\noindent
where  $Y_{lm}$ are the spherical harmonics. For the coherent two
pion photoproduction amplitudes we will use the following decomposition:

$$
F^{(\lambda)} (\vec{q}_1, \vec{q}_2; \vec{k}) =
\sum_{l_1 m_1} \sum_{l_2 m_2} {\cal F}_{l_1 m_1 l_2 m_2}^{(\lambda)}
(q_1,q_2;k) \, Y_{l_1 m_1}(\Omega_1)\, Y_{l_2 m_2} (\Omega_2)\,, \\
\eqno{(25a)}
$$

$$
V^{(\lambda)}(\vec{q}_1,\vec{q}_2;\vec{k})=
\sum_{l_1 m_1}\sum_{l_2 m_2}{\cal V}_{l_1 m_1 l_2 m_2}^{(\lambda)}
(q_1,q_2;k)\,Y_{l_1 m_1}(\Omega_1)\,Y_{l_2 m_2}(\Omega_2)\,. \\
\eqno{(25b)}
$$

\noindent
Some details for the numerical method developed for calculations
of the partial amplitudes ${\cal V}_{l_1 m_1 l_2 m_2}^{(\lambda)}$
are given in the Appendix.

Finally let us consider one useful approximation in the treatment 
of the FSI. If  in the pion nuclear Green's function

$$
\frac{1}{E(q) - E(q') + i \epsilon} = \frac{P}{E(q) - E(q')} -
i \pi \delta (E(q) - E(q'))  \\
\eqno{(26)}
$$

\noindent
we neglect the contribution from the principal value integral (first term
in Eq. (26)) and take into account that

$$
{\cal F}_l(q,q)=\frac{1}{2iq}(e^{2i\delta_l}-1) \\ \eqno{(27)}
$$

\noindent
then the FSI effects  can be taken into account in a very 
simple way in terms of the pion-nuclear elastic scattering phase 
shifts $\delta_l$, i.e.

$$
{\cal F}_{l_1 m_1 l_2 m_2}^{(\lambda)}(q_1,q_2;k)=
{\cal V}_{l_1 m_1 l_2 m_2}^{(\lambda)}(q_1,q_2;k)\,
e^{i(\delta_{l_1}+\delta_{l_2})}\,
\cos{\delta_{l_1}}\,\cos{\delta_{l_2}}\,. \\
\eqno{(28)}
$$

\noindent
A similar expression can be obtained using the K-matrix 
approach.  Therefore, we will refer to this approximation as 
"K-matrix" approximation.  Below we will see that at high 
energies $(E_{\gamma} > 600$ MeV) this approximation is good not 
only for a qualitative but also for a quantitative description 
of the FSI effects.

\section{Amplitude renormalization}
 
In the previous considerations we used the conventional approach based  
on the Impulse Approximation neglecting the influence of the nuclear 
medium on the elementary process. However, in the study of pion
nuclear elastic scattering and in the  
coherent $(\gamma,\pi^0)$ reaction\cite{Oset1,SW81,Koch1} it was 
found that the properties of the $\Delta$ resonance  in the 
nuclear medium differ from those at zero matter 
density due to many-body corrections. In the present 
paper we shall take into account the corresponding $\Delta$ 
renormalization effects using the following modification of the 
$\Delta$ propagator 

$$ 
\left[\sqrt{s}_{\Delta}-M_{\Delta}+\frac{i}{2}\Gamma\right]^{-1} 
\rightarrow \left[\sqrt{s}_{\Delta}-M_{\Delta}+ 
\frac{i}{2}{\tilde{\Gamma}}(\rho(r)) 
-\Sigma_{\Delta}(\rho(r))\right]^{-1}\,, 
\eqno{(29)} $$

\noindent
where $\rho(r)$ is the nuclear density normalized to A, 
${\tilde{\Gamma}}$ is the Pauli blocked width and $\Sigma_{\Delta}$
is the $\Delta$ selfenergy. 
The imaginary part of the
$\Delta$ selfenergy includes contribution from the two- and
three-body mechanisms of the $\Delta-h$ interaction and the real part
takes into account the Landau-Migdal force 
$g_{\Delta}'(f^*/m_{\pi})^2\vec{S}\cdot\vec{S}^{\dagger}\,
\vec{T}\cdot\vec{T}^{\dagger}$, which leads to the following 
renormalization effects~\cite{Carrasco}

$$
  Re \Sigma_{\Delta}\rightarrow Re \Sigma_{\Delta}' +
\frac{4}{9}(\frac{f^*}{m_{\pi}})^2\,g_{\Delta}'\,\rho(r)
\eqno{(30)}
$$
with  $g_{\Delta}'$=0.55. Parametrizations for 
${\tilde{\Gamma}}_{\Delta}$, $Re \Sigma'_\Delta$
and $Im \Sigma_\Delta$ can be found in Refs.~\cite{Carrasco,Garcia}.
Our analysis and calculations in 
Ref.~\cite{Carrasco} show that such renormalization of the  
$\Delta$ propagator provides a good 
description of the coherent $(\gamma,\pi^0)$ reaction on $^{12}C$. This
circumstance gives us confidence in the study of the medium effects  in the
coherent two pion photoproduction.

\section{Results and Discussion}

{\underline {\it Symmetry properties}}.
In the case of the two pion photoproduction, in contrast to the 
single pion photoproduction, we have more independent 
kinematical variables in the final state and, therefore, more 
possibilities for the study of the mechanisms of this reaction. 
One of the interesting kinematical regions corresponds to the 
case when two pions  propagate together, i.e. in the same 
direction and with equal energies.  In Fig. 3 we present results 
of the PWIA analysis at $\theta_1=\theta_2=15^0$ and at  
$\phi=\phi_1-\phi_2= 0, 90^0$ and $180^0$. In the case of 
$\phi=0$ the pions propagate in the same direction.  In the case 
of $\phi=180^0$ they are in the same plane with the initial 
photon and in opposite sides of it.  

One of the main conclusions which follows from the results 
depicted in Fig. 3a is that the probability of propagation of 
$\pi^+$- and $\pi^-$-mesons together  (which corresponds to the 
middle of the energy distribution at $\phi=0$) is very small ( 
about 0.1 nb). They prefer to propagate in one plane with the 
photon with $\phi=180^0$.  In the $(\gamma,\pi^0\pi^0)$ channel 
(see Fig. 3b) the situation is reversed. The two $\pi^0$ mesons 
prefer to propagate together.  However, the cross section in 
this kinematical region is about three orders of magnitude 
smaller than in the $(\gamma,\pi^-\pi^+)$ channel when the 
charged pions go out with $\phi=180^0$.  In order to understand 
such behaviour let us recall the symmetry properties considered 
in Section 1.  From Eq. (10) we have

$$
\frac{d\sigma}{dT_1 d\Omega_1 d\Omega_2}(\phi=0)=
\frac{S_B}{(2\pi)^5}\,\frac{q_1 q_2 }{16 E_{\gamma}}\, 
\mid F_1 \sin\theta_1 +  F_2 \sin\theta_2 \mid^2\,, \\
\eqno{(31a)}
$$

$$
\frac{d\sigma}{dT_1 d\Omega_1 d\Omega_2}(\phi=180^0)=
\frac{S_B}{(2\pi)^5}\,\frac{q_1 q_2 }{16 E_{\gamma}}\, 
\mid F_1 \sin\theta_1 -  F_2 \sin\theta_2 \mid^2\,. \\
\eqno{(31b)}
$$

In the coherent $(\gamma,\pi^0\pi^0)$ reaction on the 
isospin-zero nuclei only isoscalar photons give the contribution 
in the cross section and they create two pions in the symmetric 
state with isospin $T=0$.  Therefore, as it follows from Eq. 
(5), $F_{1} (\vec{q}_1,\vec{q}_2;\vec{k}) = 
F_{2} (\vec{q}_2, \vec{q}_1; \vec{k})$ and the maximum cross 
section is in the region where $\vec{q}_1= \vec{q}_2$ (two pions 
propagate together).

In the  $(\gamma,\pi^-\pi^+)$  channel both isoscalar and 
isovector photons contribute in the cross section, but from the 
analysis of the $\pi^0\pi^0$ photoproduction it follows that the 
role of the isoscalar photons is very small. Therefore, the 
dominant contribution in the $\pi^-\pi^+$ channel comes from the 
isovector photons which produce pions in the antisymmetric state 
with isospin $T=1$.  Since in this case (see Eq. (4)) 
$F_{1}(\vec{q}_1,\vec{q}_2;\vec{k})= 
-F_{2}(\vec{q}_2,\vec{q}_1;\vec{k})$ the differential cross 
section has a dip minimum at $\vec{q}_1=\vec{q}_2$ region which 
is filled only by small contributions from the isoscalar 
photons. For the same reason the maximum cross section is found 
when $\phi=180^0$.

Note that this interesting feature of the coherent two pion 
photoproduction is based only on the isospin symmetry properties 
and it does not depend on the mechanisms of the process or FSI 
effects.  If an experimental study would find serious deviations 
from our predictions (for example, nonzero cross section in the 
kinematical region where energies and polar angles of the two 
$\pi^0$ mesons are equal and their relative azimuthal angle 
$\phi=180^0$ ) this would be an indication of isospin violation 
effects.

{\underline{\it The study of the elementary amplitude}}. 
Another feature is related directly to the dynamical aspects of 
the reaction.  In Fig. 4 the energy distributions of the $\pi^-$ 
mesons calculated in the PWIA approach at $E_{\gamma}$=450, 600 
and 750 MeV are depicted.  Here we analyse the contributions 
coming from different parts of the elementary amplitude (see 
also Fig. 2):  Born terms (diagrams (a)-(h)), $\Delta$ isobar 
terms (diagrams (i)-(o)) and contributions from  $D_{13}(1520)$ 
and $P_{11}(1440)$ baryon resonances (diagrams (p)-(t)). In the 
$(\gamma,\pi^-\pi^+)$ channel the contribution of the Born terms 
is very small (about 1-2\%) in a large  photon energy region. 
The dominant contribution is related to the excitation of the 
$\Delta$ isobar (dashed curves in Fig. 4).  The contributions 
coming from the $P_{11}(1440)$ and $D_{13}(1520)$ resonances 
tend to cancel each other. In the $(\gamma,\pi^0\pi^0)$ channel 
the situation is similar with only one exception: the Born terms 
at low energies $(E_{\gamma}<450$MeV) are also important.

Finally, note that the energy  distribution in the 
$(\gamma,\pi^0\pi^0)$ channel is exactly symmetrical relative to 
point where the energies of the pions are equal $(T_{{\pi}_1}= 
T_{{\pi}_2})$. However in the $(\gamma,\pi^-\pi^+)$ channel 
this symmetry is slightly destroyed due to the interference of 
the small contribution from the isoscalar photons with the 
dominant contributions from isovector photons.  

{\underline {\it Final State Interaction effects}}.  
The treatment of the FSI effects in the case of the two pion 
photoproduction is more complicated in contrast to the single 
pion photoproduction case.  From the study of the last 
one~\cite{SW81,Koch1,Boffi,Aset,Carrasco} it is well known that FSI 
increases the differential cross section at low energies 
($T_{\pi}<100$) MeV  and reduces it in the $\Delta$ resonance 
region (in about a factor two). 

However, in the two pion photoproduction if one pion has low 
energy the second pion can be produced in the $\Delta$ resonance 
region. Therefore, the FSI of one pion can be canceled by the 
FSI of the second one. This peculiarity of the FSI is 
illustrated in Fig. 5 for the case of $E_{\gamma}=450$ MeV. If 
only one  pion is interacting with the residual nucleus, then 
the FSI leads to the enhancement of the cross section in the 
low-energy part of the phase space and to the reduction in the 
high-energy part (compare the dotted curve with the dashed or 
with the dash-dotted curves). In the full calculations (solid 
curve) the FSI increase the cross section only in the region 
where both pions are at relatively low energies ( in the middle 
of the phase space).  Note that due to the Coulomb interaction 
the  FSI effects are larger for low-energy $\pi^-$ mesons than 
for the low-energy $\pi^+$ mesons. This produces an additional 
asymmetry in the energy distribution relative the point where 
the pions have equal energy. 

{\underline {\it The off-shell effects}}.
The FSI contributions contain one ingredient which is not well 
defined. This is the propagation of the pions in the off-shell 
region where $\omega\neq\sqrt{m_{\pi}^2+\vec{q}\,^2}$. In terms 
of the pion-nuclear Green's function (26) it corresponds to 
contributions coming from the principal value integral. In this 
part we have to define the elementary amplitude $t^{(\lambda)}$ 
also for the photoproduction of off-shell pions.

On the other hand the pion-nuclear scattering amplitude $F_{\pi 
A}(\vec{q'},\vec{q})$ which is normally extracted  from the 
pion-nuclear elastic scattering data is well defined only in the 
on-shell region (at $q'=q)$. Its off-shell behavior, which 
specifies the dynamics of the pion-nuclear system in the 
interacting region (inside the nucleus), depends on the  
extrapolation of the elementary pion-nucleon amplitude to the 
off-shell region.
The off-shell dependence of these amplitudes can have 
repercussion in production processes. In coordinate space 
this could be stated as saying that optical
potentials which provide the same scattering properties can lead to 
different results in the production  process.

To study the sensitivity of the two pion photoproduction to 
these ingredients of the theory we shall use the standard 
prescription for the extrapolation of the elementary amplitude 
in the off-shell region i.e.

$$
t^{(\lambda)}(\vec{q}_1,\vec{q}_2;\vec{k})\rightarrow
t^{(\lambda)}(\vec{q}_1,\vec{q}_2;\vec{k})\,F(q_1)\,F(q_2)\,, \\
\eqno{(32)}
$$

\noindent
where for the form factor $F(q)$ we  use monopole type expression

$$
F(q)=\frac{\Lambda^2-m_{\pi}^2}{\Lambda^2-\omega^2+\vec{q}\,^2}\, . \\
\eqno{(33)}
$$

\noindent
Note that this procedure can be interpreted also as a modification of 
the pion-nuclear scattering amplitude $F_{\pi A}$ in the 
off-shell region without changing its on-shell values (in other 
words, to use a phase-equivalent optical potentials which 
provide different behavior of the pion wave function in the 
interaction region and the same asymptotic). 

In Fig. 6, for the photon energy 450 MeV, we illustrate the 
sensitivity of the FSI contributions to the different values of 
the cut-off parameter $\Lambda$=500, 1000 and 1500 MeV. We see 
that the DWIA results caused by different off-shell 
extrapolations can differ by about 50\%.  However, with 
increasing photon energy the contribution from the off-shell 
region reduces (compare solid and dashed curves). Starting from 
$E_{\gamma}$=600 MeV the simple "K-matrix" approximation (see 
Eq. (28)), where we deal only with on-shell pions, becomes a 
reliable approach to account for FSI effects. In this region 
they are very important and can be described only with 
pion-nuclear scattering phase shifts. 

{\underline {\it Delta renormalization effects}}.  One of the 
important problem which can be studied in the coherent two pion 
photoproduction is the modification of the elementary operator 
in the nuclear medium. From the analysis of the 
elementary amplitude, we have seen above that the main one-body 
mechanism is related to the excitation of the $\Delta$-isobar. 
This finding allows us to do reliable investigations of the 
$\Delta$-renormalization effects, as it was done in the 
coherent single $\pi^0$-photoproduction~\cite{SW81, Koch1,Carrasco}.

In Fig. 7 we compare the conventional DWIA 
calculations presented above with the results obtained including 
the renormalization 
of the $\Delta$-propagators in accordance with Eqs. (29,30). Note 
that the corresponding medium effects  are different in the 
different kinematical regions. They are maximal in the region 
where one of the pions is in the $\Delta$-resonance region 
(around $T_{\pi}=$180 MeV) and become less relevant at lower and higher 
energies. Due to this fact, the renormalization of the 
$\Delta$-propagators changes not only the absolute value of the 
differential cross section, but due to the presence of the two 
pions it also modifies the shape of the energy distribution.  
The most sizeable effects are seen at 
$E_{\gamma}$=750 MeV. Thus we have found that medium effects in 
the coherent two pion photoproduction are very important and can 
be observed experimentally.

{\underline {\it Angular correlation and total cross section}}.
In the next Fig. 8 the dependence on the pion polar angle 
$\theta$ (angular correlation) is depicted both for the 
$\pi^-\pi^+$ and $\pi^0\pi^0$ channels. From the discussions of 
the symmetry properties it follows that the differential cross 
section reaches the maximal value in the middle of the phase 
space at $\phi=0$ in the $\pi^0\pi^0$  and at $\phi=180^0$ in 
the $\pi^-\pi^+$ channels (see also Fig. 3). We have done the 
calculations for these kinematical regions assuming that 
$\theta_1=\theta_2$.

One of the peculiarities which follows from the general symmetry 
properties is that the angular correlation function vanishes at 
$\theta_1=\theta_2=$ 0 or $180^0$ due to the $\sin\theta_{1(2)}$ 
dependence of the differential cross section. Furthermore, in 
the $\pi^0\pi^0$ channel we are already in the region where the 
nuclear form factor has a minimum. This is because at $\phi=0$ 
the  transferred momentum is larger than at $\phi=180^0$ (see 
Eq. (17)).  

The energy dependence of the total cross section (integrated 
over the whole phase space) is depicted in Fig. 9.  Here we 
compare  the PWIA , DWIA and results obtained with $\Delta$ 
renormalization. At $350<E_{\gamma}<450$ 
MeV  the FSI can increase the cross section up to a factor two.  
Note that such enhancement is smaller than what was found in the 
case of inclusive double photoproduction~\cite{Jose2}. The 
reason must be traced to the fact that, in contrast to the 
coherent photoproduction, in the inclusive reaction, due to the 
energy transfer to the residual nucleus, both pions can be in 
the low-energy region where the enhancement from the $\pi A$ 
interaction shows up. At $E_{\gamma}>500$ MeV FSI reduces the total cross
section up to a factor 4-5. The further reduction comes from the 
$\Delta$ renormalization effects.

Finally, let us make one comment about the $\pi^0\pi^0$ channel. 
We have found that in this channel the cross section is about 
1000 times smaller than  in the $\pi^-\pi^+$ channel. In such a 
situation we expect that in the $\pi^0\pi^0$ channel the 
contribution of the pion rescattering on the different nucleons 
with charge exchange could be important.  Note that similar 
effects have been  found in the single $\pi^0$ photoproduction 
on the deuteron \cite{Koch2,Bosted}. Moreover, recently in the  
$(\gamma,\pi^0\pi^0)$ reaction on the proton \cite{Bern2} at 
threshold region it was shown that such rescattering effects (on 
the same nucleon) are important. The role of the 
coupling to the charge-exchange channel mentioned above is beyond the DWIA 
approach and could be the subject of further investigations.  We 
expect that due to this mechanism, the $\pi^0\pi^0$ channel 
could be appreciably enhanced.  

On the other hand the fact that the two $\pi^0$ mesons come 
mostly together and in a $T=0$ state could be used to 
investigate possible renormalization effects of such pionic 
state, which have been suggested in 
Refs. \cite{Chanfray1,Chanfray2}.

\section{Conclusions}

Using a recently developed model for the two pion 
photoproduction on the nucleons based on Effective Lagrangians 
we have studied the coherent $(\gamma,\pi^-\pi^+)$ and 
$(\gamma,\pi^0\pi^0)$ reactions in $^{12}C$.  The coherence of 
these reactions and its study in isospin-zero nuclei has allowed 
us to see interesting effects tied to the bosonic symmetry of 
the pions which do not depend on the reaction mechanisms. 

We have found that in the $(\gamma,\pi^-\pi^+)$ channel the 
dominant contribution comes from the isovector part of the 
electromagnetic current.  In this case the probability of 
production of $\pi^-$- and $\pi^+$-mesons propagating together 
is very small.  On the other hand, only the isoscalar 
electromagnetic current gives the contribution in the 
$(\gamma,\pi^0\pi^0)$ channel. In this case, due to isospin 
symmetry arguments, two $\pi^0$-mesons prefer to propagate 
together. On the other hand, the kinematical region where the energies 
and polar angles of the two $\pi^0$ mesons are equal and their 
relative azimuthal angle $\phi = 180^0$ becomes forbidden if the 
total isospin is a good quantum number.  

The main one-body mechanism in both channels is related to the 
excitation of the $\Delta$ isobar.  This finding could be 
used in future investigations of the renormalization of the 
$\Delta$ propagator in the nuclear medium, as it was 
done in the coherent $(\gamma,\pi^0)$ reaction. Many-body 
effects caused by pion rescattering (or FSI) and $\Delta$ 
renormalization are  very 
important. We have found that near threshold FSI can increase 
the cross section in about a factor two.  At $E_{\gamma}>600$ 
MeV it reduces the cross section in about a factor 4-5.  In 
this region the rescattering of the on-shell pions is dominant 
and with good accuracy the FSI effects can be described  using 
simple "K-matrix" approximation. The 
$\Delta$ renormalization leads to an important reduction of the
cross section and it modifies the shape of the energy distribution
of the pions.

\vspace{0.2cm}

{\bf Acknowledgments}:
This work is partially supported by the CICYT contract no. 
AEN-96-1719.  We would like to thank J. A. Gomez Tejedor for 
multiple discussions concerning the elementary $(\gamma,\pi\pi)$ 
model, S. K. Singh and M. J. Vicente-Vacas for the discussions 
of the isospin symmetry.  One of us (S.S.K.) wants to 
acknowledge support from the Ministerio de Educacion y Ciencia 
in his sabbatical stay at the University of Valencia.

\vspace{0.5cm}
\noindent
{\Large\bf Appendix}
\vspace{0.5cm}

In this Appendix we present some details for the calculations of 
the partial amplitudes  ${\cal V}_{l_1 m_1 l_2 
m_2}^{(\lambda)}$. In accordance with Eq. (25) they are defined 
as

$$
{\cal V}_{l_1 m_1 l_2 m_2}^{(\lambda)}(q_1,q_2;k)=
\int V^{(\lambda)}(\vec{q}_1,\vec{q}_2;\vec{k})\,
Y_{l_1 m_1}^*(\Omega_1)\,Y_{l_2 m_2}^*(\Omega_2)\,d\Omega_1\,d\Omega_2\,.
\\
\eqno{(A1)}
$$

In order to perform the integration over the pions azimuthal 
angles $\phi_1$ and $\phi_2$ we explore the fact that the 
$V_{1(2)}$ amplitudes depend on the relative azimuthal angle 
$\phi=\phi_1-\phi_2$.  Therefore, we can introduce a fastly 
convergent expansion

$$
V_{1(2)}(\vec{q}_1,\vec{q}_2;\vec{k})=\sum_m\,
V_{1(2)}^{(m)}(q_1,q_2,x_1,x_2;k)\,e^{im\phi}\, \\
\eqno{(A2)}
$$

\noindent
where $x_{1(2)}=\cos{\theta_{1(2)}}$. The inverse expression is given by

$$
V_{1(2)}^{(m)}(q_1,q_2,x_1,x_2;k)=\frac{1}{2\pi}\,\int_0^{2\pi}
V_{1(2)}(\vec{q}_1,\vec{q}_2;\vec{k})\,e^{-im\phi}\,d\phi\,. \\
\eqno({A3)}
$$

Using Chebyshev quadrature for numerical integration in Eq. (A3)
for the $V_{1(2)}^{(m)}$ amplitude we obtain  

$$
V_{1(2)}^{(m)}(q_1,q_2,x_1,x_2;k)=\frac{1}{N}\sum_{i=1}^{N}\,
V_{1(2)}(\vec{q}_1\!^i,\vec{q}_2\!^i;\vec{k})\,Re[e^{-im\phi_i}]\,, \\
\eqno{(A4)}
$$

\noindent
where $\cos{\phi_i}=\cos\left[\frac{(2i-1)\pi}{N}\right]$ and 
$\vec{q}_{1(2)}\!\!\!\!\!\!^i\quad$ are the corresponding pion momenta.

After the integration over $\phi_1$ and $\phi_2$ angles in Eq. 
(A1) the partial amplitudes ${\cal V}_{l_1 m_1 l_2 
m_2}^{(\lambda)}$ can be expressed in terms only of the 
$V_{1(2)}^{(m)}$ amplitudes:

$$
\begin{array}{ll}
{\cal V}_{l_1 m_1 l_2 m_2}^{(\lambda)}  = &
-\lambda\,(2\pi)^{3/2}\,\int_{-1}^{1}\int_{-1}^{1}dx_1\,dx_2\,
\left[ V_1^{(m_1-\lambda)}\sin\theta_1+ V_2^{(m_1)}\sin\theta_2\right] \\ [2ex]
& \times Y_{l_1 m_1}(x_1,\phi_1=0)\,Y_{l_2 m_2}(x_2,\phi_2=0)\,, 
\hspace{5.2cm} (A5)
\end{array}
$$

\noindent
where $m_2=-m_1+\lambda$. For the numerical integration in Eq. (A5) 
we use the Gaussian quadrature.

\newpage

\noindent
{\Large\bf Figure captions}

\vspace{0.5cm}
\noindent
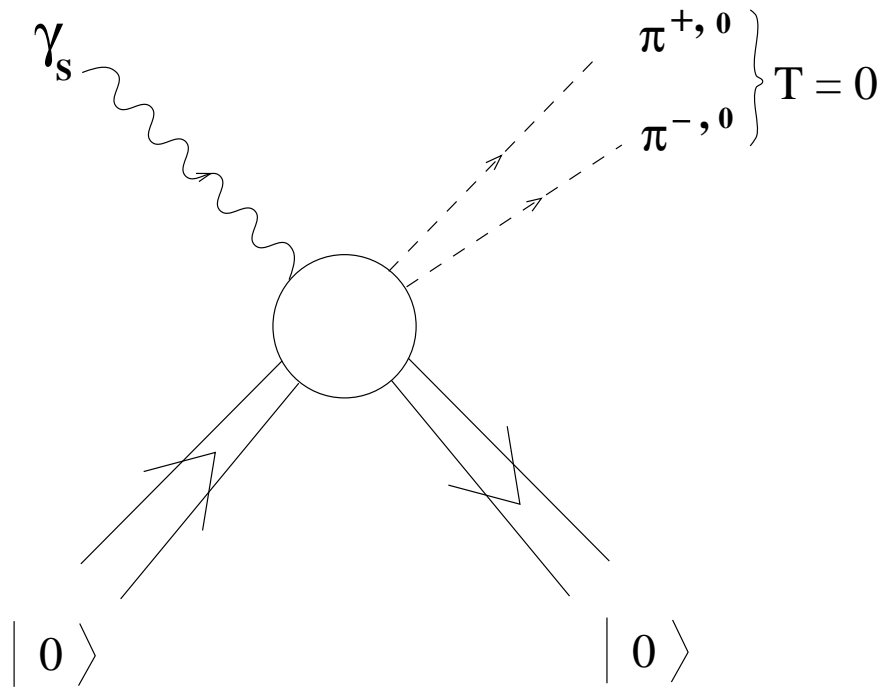
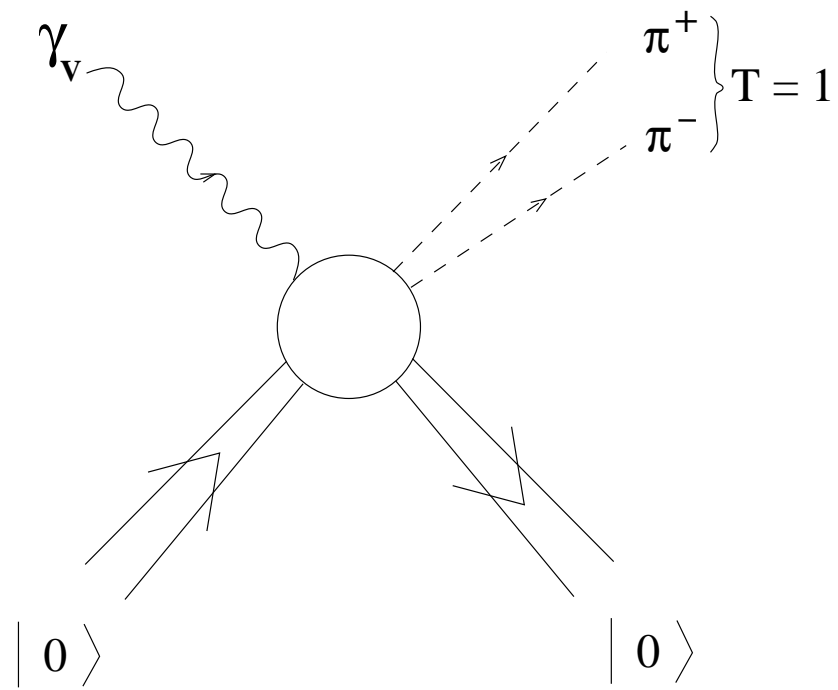
{\bf Fig. 1}.  Isospin selection rules for the coherent two pion 
photoproduction on the isospin zero targets.

\vspace{0.5cm}
\noindent
{\bf Fig. 2}.  Feynman diagrams of the model of Ref.\cite{Jose1} 
for the reaction $\gamma N\rightarrow N\pi\pi$. 

\vspace{0.5cm}
\noindent
{\bf Fig. 3}.  Energy spectra of the $\pi^-${\bf (a)} and $\pi^0$ 
{\bf (b)} mesons in the $(\gamma,\pi^-\pi^+)$ and 
$(\gamma,\pi^-\pi^+)$ channels, respectively, at 
$E_{\gamma}=600$ MeV and $\theta_1=\theta_2=15^0$. Dashed, 
dash-dotted and solid curves are the PWIA calculations at 
$\phi=0, 90$ and $180^0$, respectively.

\vspace{0.5cm}
\noindent
{\bf Fig. 4}.  Energy spectra ( integrated over pions solid angles) 
of the $\pi^-${\bf (a)} and $\pi^0$ {\bf (b)} mesons in the 
$(\gamma,\pi^-\pi^+)$ and $(\gamma,\pi^0\pi^0)$ channels, 
respectively.  at $E_{\gamma}=450$, 600 and 750 MeV. Dashed, 
dotted and solid curves are the PWIA calculations with $\Delta$, 
$\Delta + D_{13}$, and $\Delta + Born +P_{11}+D_{13}$ terms, 
respectively.

\vspace{0.5cm}
\noindent
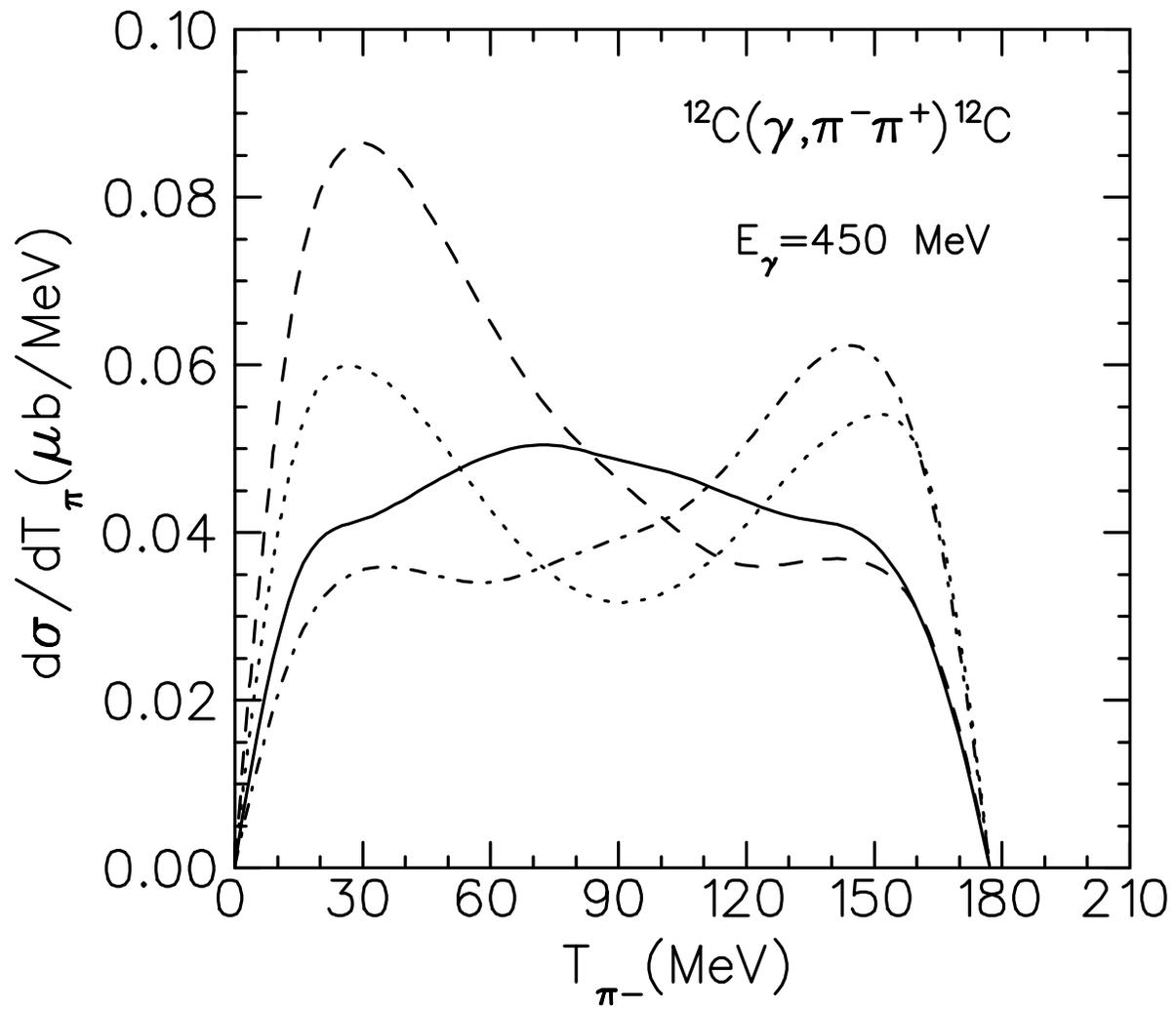
{\bf Fig. 5}.  FSI effects in the  $(\gamma,\pi^-\pi^+)$ channel at 
$E_{\gamma}=450$ MeV. The dotted curve is the PWIA result. The 
dashed and dash-dotted curves are the DWIA calculations when 
only $\pi^-$- or $\pi^+$-mesons  are interacting with the 
residual nucleus, respectively.  The solid curve is the full 
DWIA results obtained with $\Lambda$=1 GeV.

\vspace{0.5cm}
\noindent
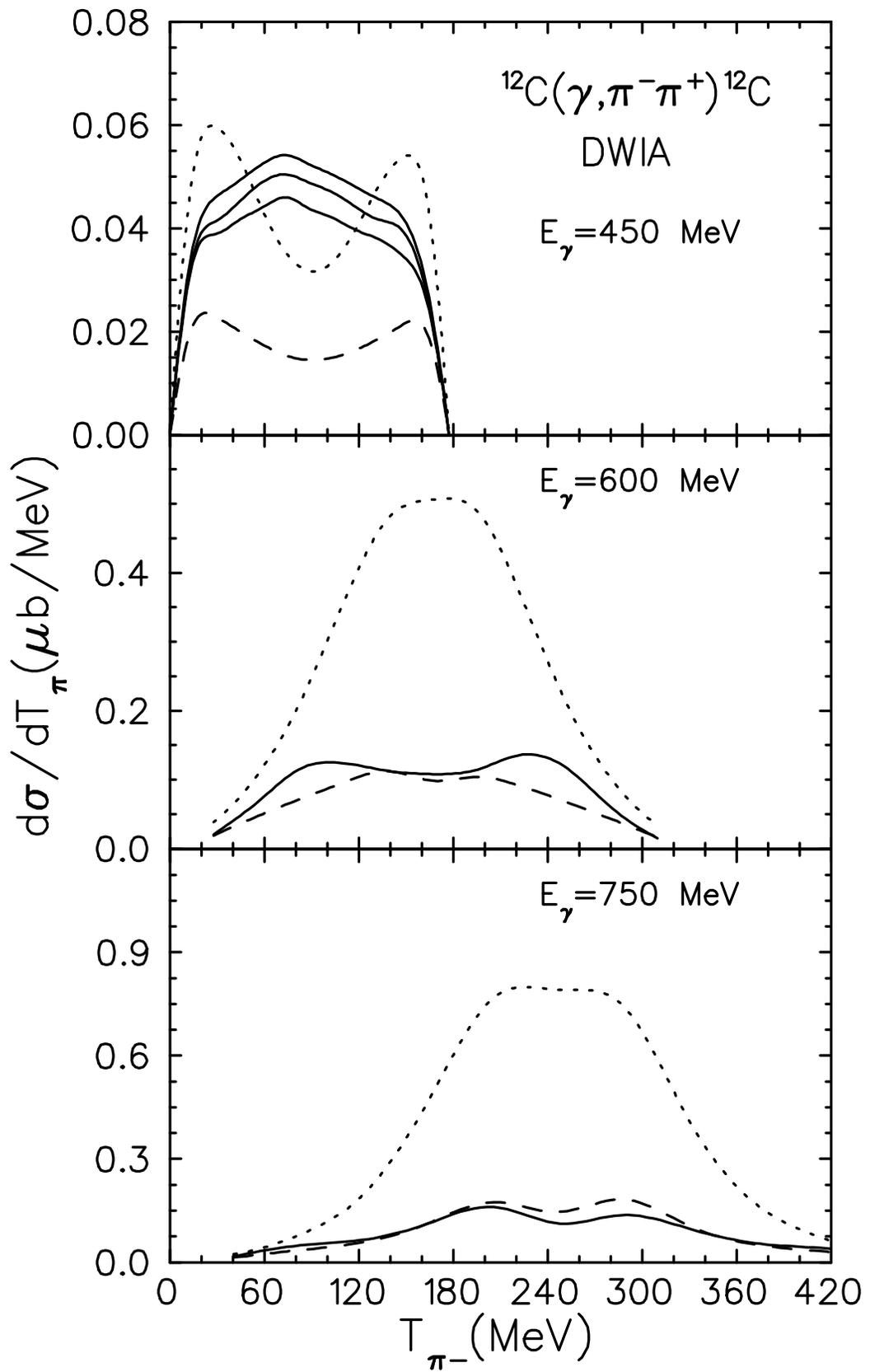
{\bf Fig. 6}.  Off-shell effects  in the FSI. The dotted curves are 
the PWIA results.  The dashed curves are the results obtained 
using the "K-matrix" approximation ( see Eq. (28)). At 
$E_{\gamma}=450$ MeV the upper, middle and lower solid curves 
are the full DWIA results with $\Lambda=1.5$, 1.0 and 0.5 GeV, 
respectively. At $E_{\gamma}$=600 and 750 MeV the solid curves 
are full DWIA results with $\Lambda=$ 1.0 GeV. 

\vspace{0.5cm}
\noindent
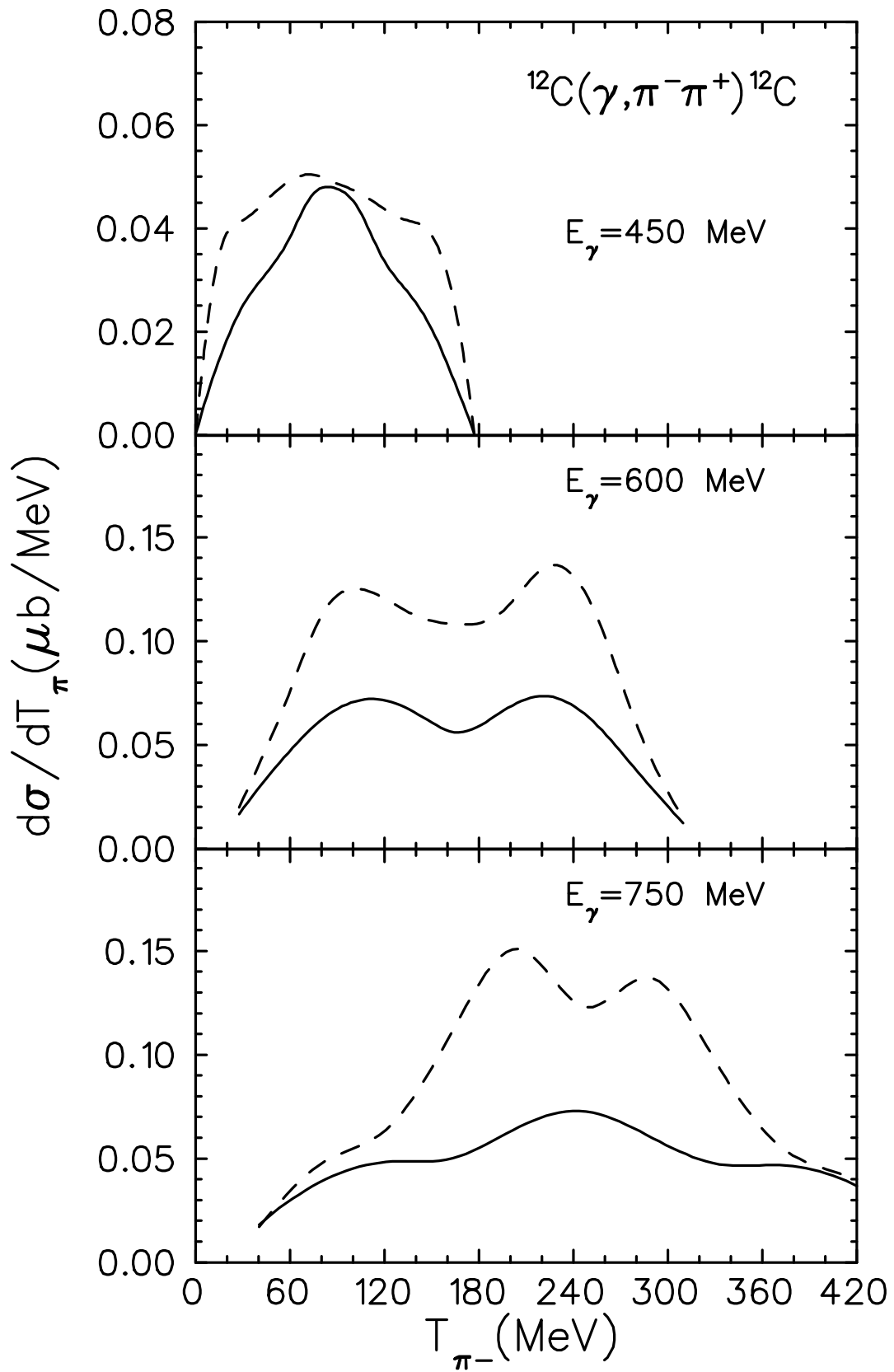
{\bf Fig. 7}.  $\Delta$ renormalization effects. The dashed curves are 
the DWIA (with $\Lambda=1$ GeV) results.  
The solid curves are the results obtained 
with renormalized $\Delta$ propagator ( see Eqs. (29,30)).

\vspace{0.5cm}
\noindent
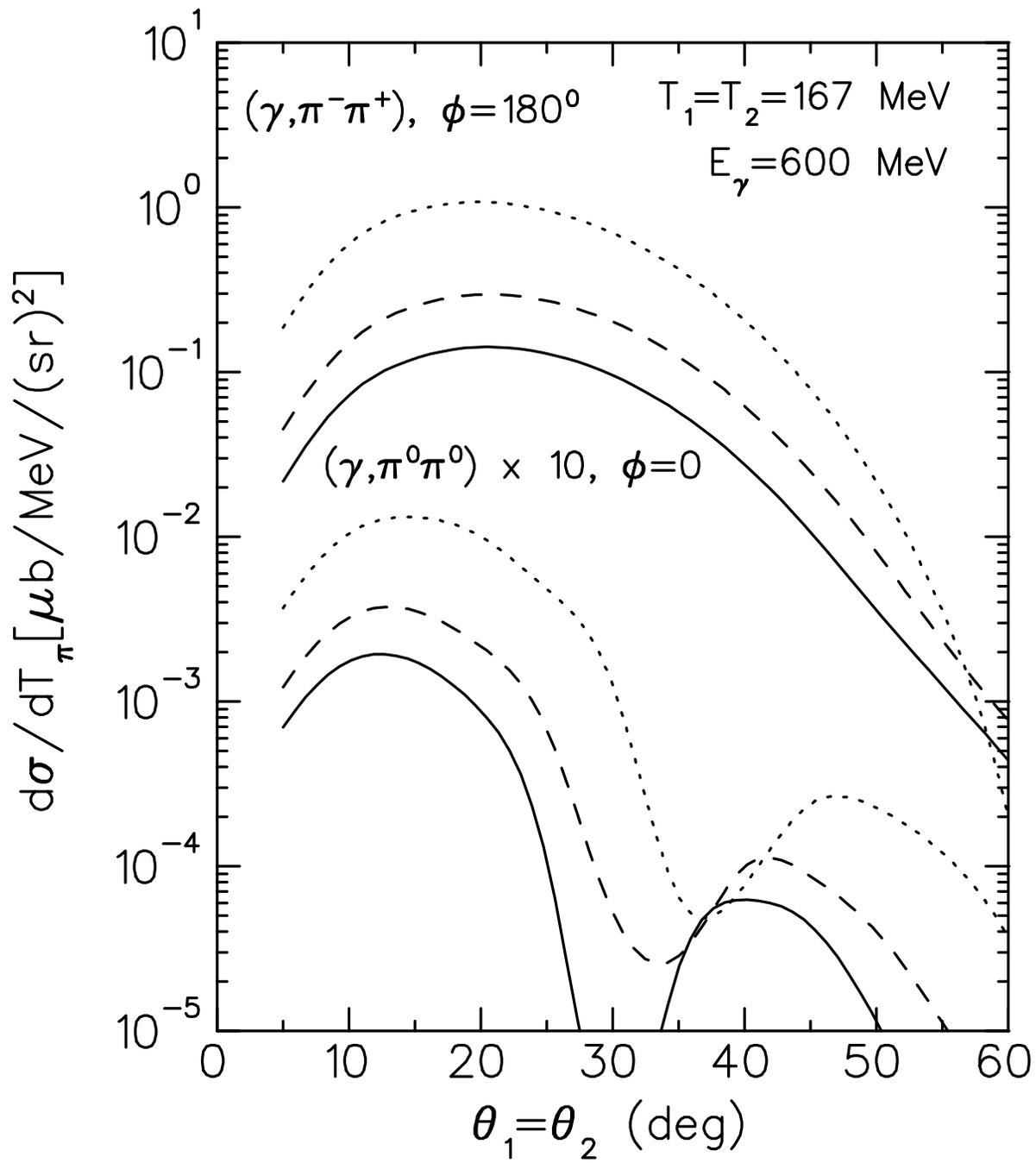
{\bf Fig. 8}.  Angular distribution of the $\pi^-$-mesons in the 
$(\gamma,\pi^-\pi^+)$ reaction at $\phi=180^0$ and of the 
$\pi^0$-mesons in the $(\gamma,\pi^0\pi^0)$ reaction at $\phi=0$  
in the middle of the phase space. The dotted and dashed curves 
are the PWIA and full DWIA (with $\Lambda=1$ GeV) calculations, 
respectively.
The solid curves are the results obtained 
with renormalized $\Delta$ propagator.

\vspace{0.5cm}
\noindent
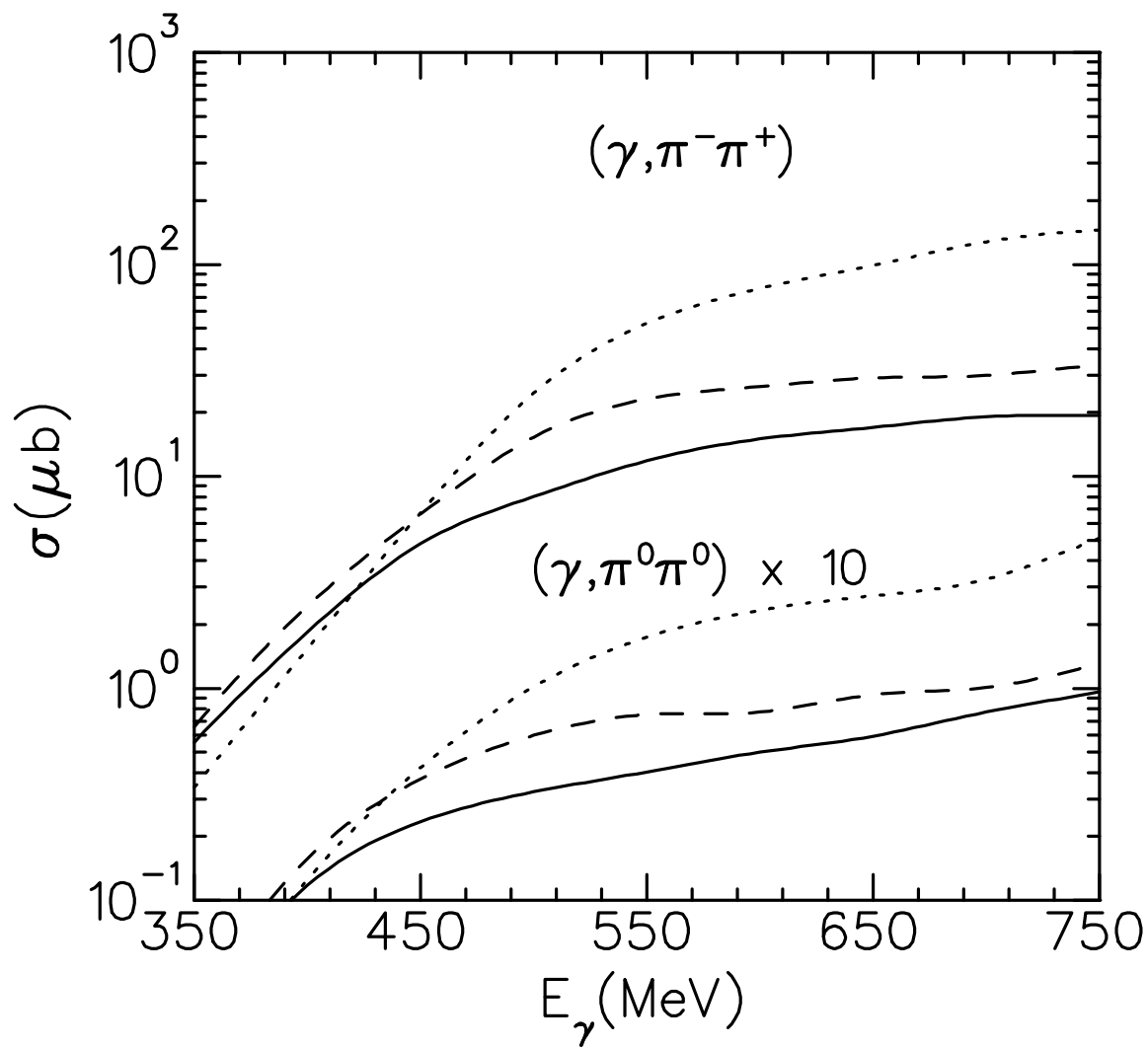
{\bf Fig. 9}.  Cross section for the coherent $(\gamma,\pi^-\pi^+)$ 
and $(\gamma,\pi^0\pi^0)$ reactions on $^{12}C$ as a function of 
photon energy. Notations of the curves are the same as in Fig. 8.

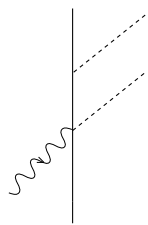
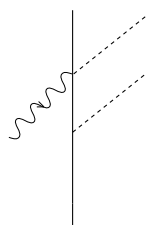
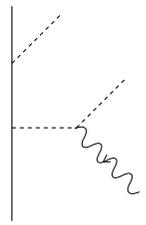
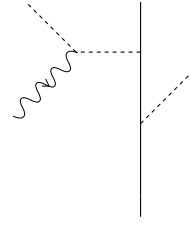
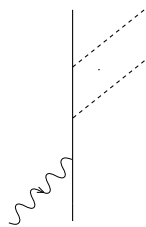
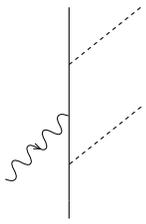
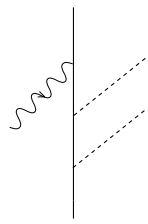
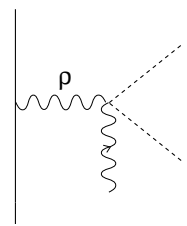
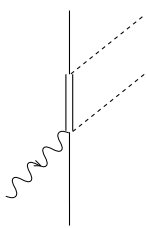
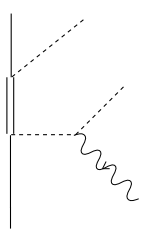
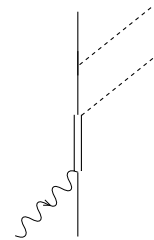
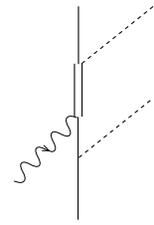
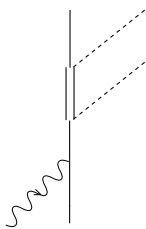
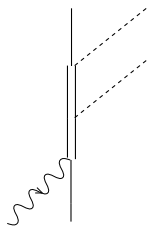
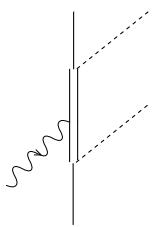
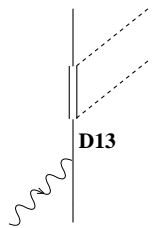
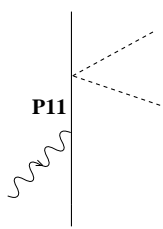
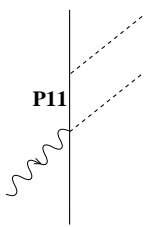
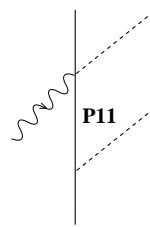
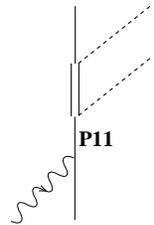
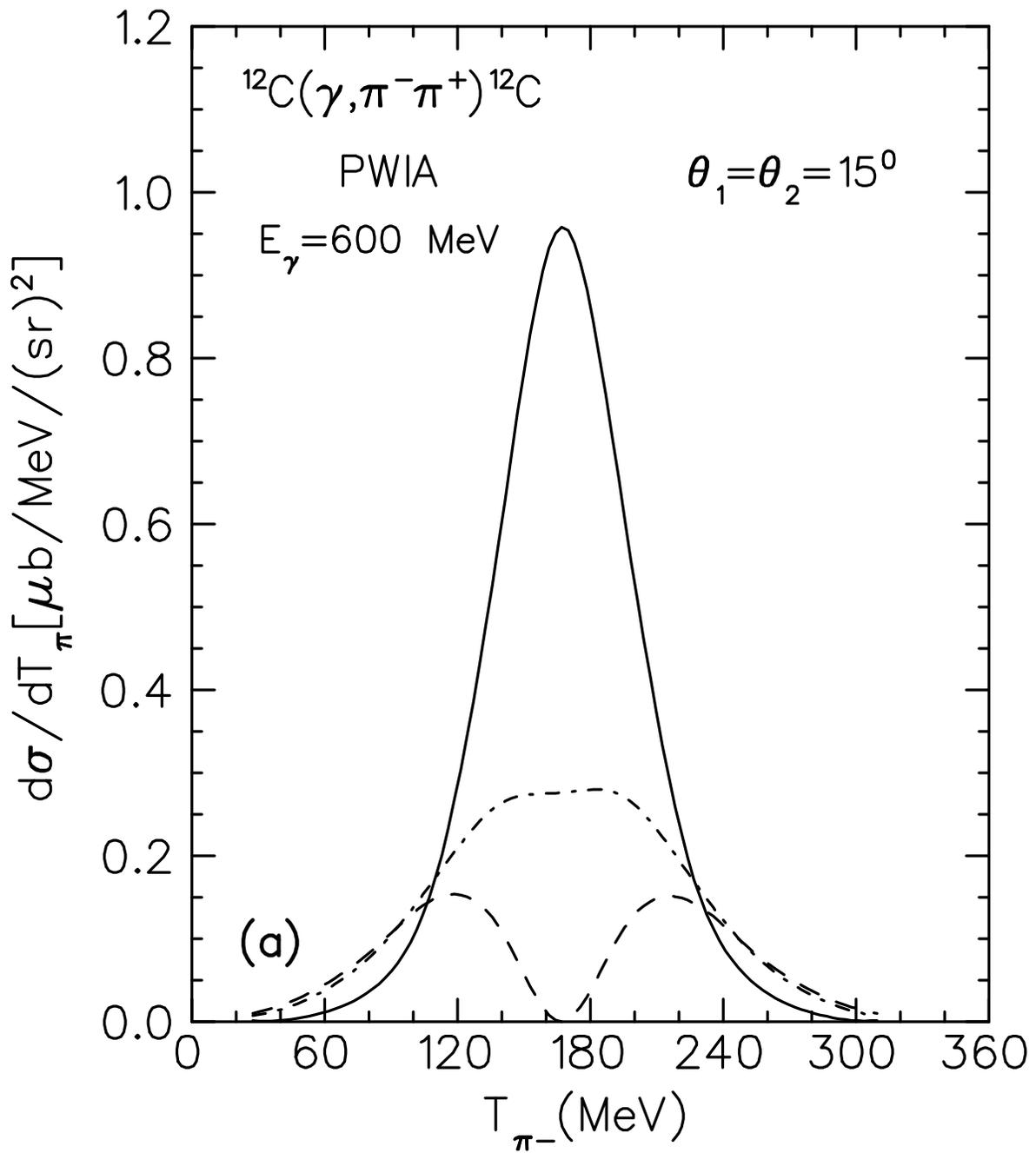
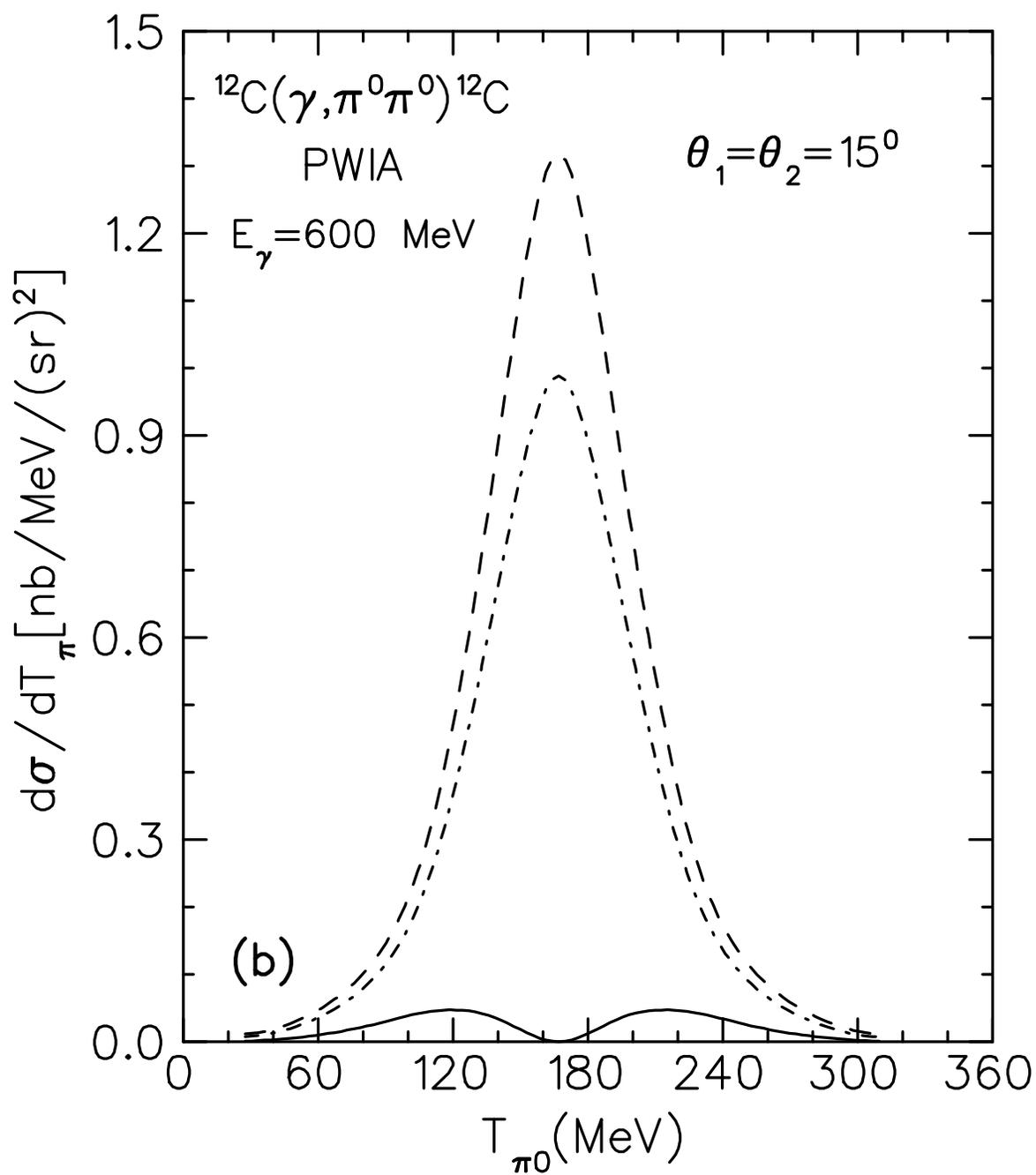
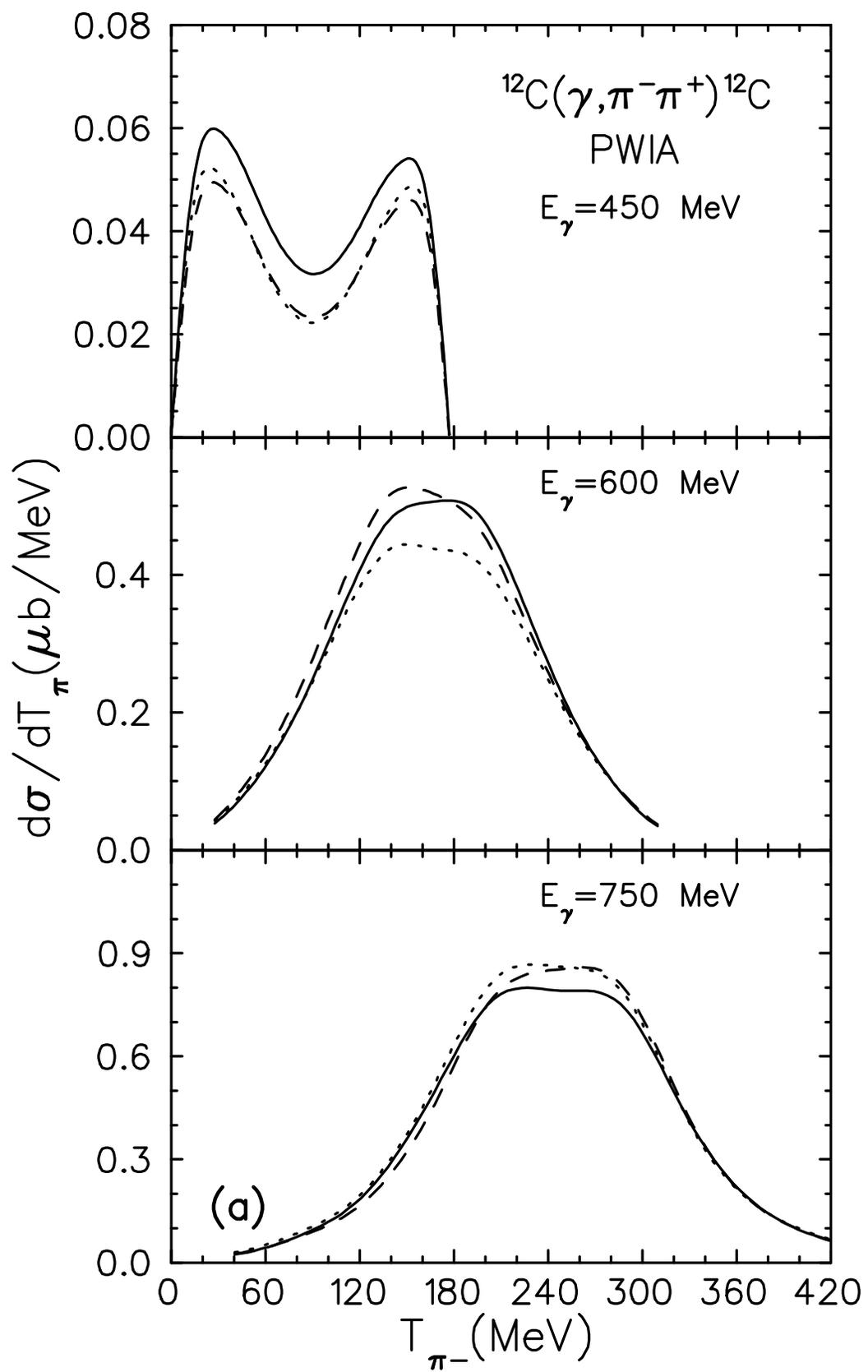
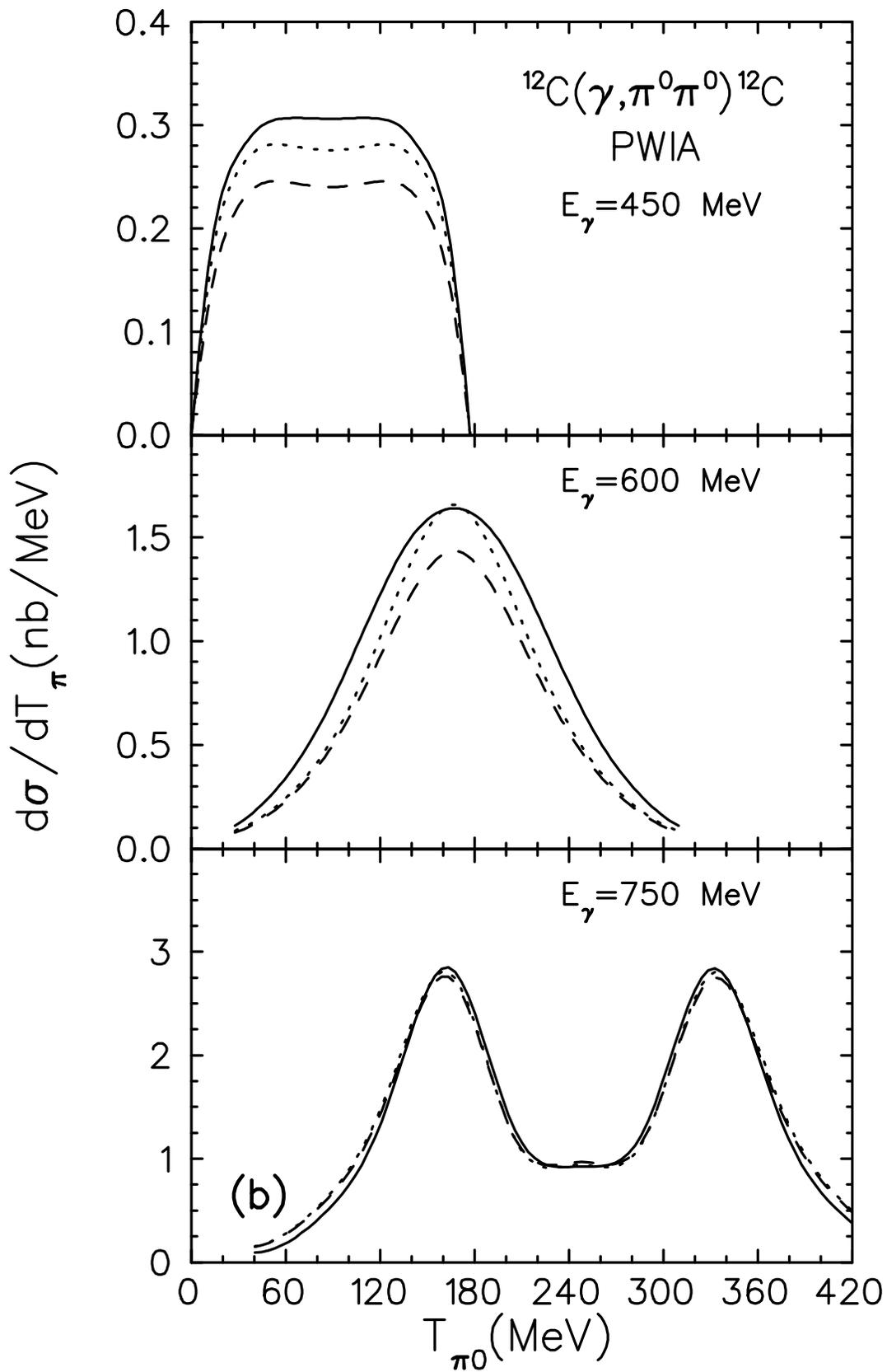
\end{document}